\documentclass[aps,twocolumn,superscriptaddress,nofootinbib,prr]{revtex4-1}
\usepackage[pdftex]{graphicx}
\usepackage{latexsym,amsmath,verbatim}
\usepackage{xcolor}
\usepackage[hidelinks,unicode=true]{hyperref}
\hypersetup{colorlinks=true,linkcolor=blue,urlcolor=blue,citecolor=blue,pdfhighlight=/N
}
\usepackage{lipsum}
\usepackage[english]{babel}
\usepackage{microtype}
\usepackage[normalem]{ulem}

\newcommand{\av}[1]{\left\langle {#1} \right\rangle}
\newcommand{\kmax}{k_\mathrm{max}}
\newcommand{\be}{\begin{equation}}
\newcommand{\ee}{\end{equation}}

\begin{document}
\title{Influential spreaders for recurrent epidemics on networks}

\author{Ga\"el Poux-M\'edard}
\affiliation{\'Ecole Normale Sup\'erieure (ENS) de Lyon,
15 parvis Ren\'e Descartes, 69342 Lyon, France}
\affiliation{Istituto dei Sistemi Complessi (ISC-CNR), Via dei Taurini 19,
I-00185 Roma, Italy}
\affiliation{Laboratoire ERIC, Universit\'e Lumi\`ere Lyon 2, 
5 avenue Pierre Mend\`es France, 69676 Bron, France}

\author{Romualdo Pastor-Satorras}

\affiliation{Departament de F\'{\i}sica, Universitat Polit\`ecnica de
Catalunya, Campus Nord B4, 08034 Barcelona, Spain}

\author{Claudio Castellano}
\affiliation{Istituto dei Sistemi Complessi (ISC-CNR), Via dei Taurini 19,
I-00185 Roma, Italy}
%\affiliation{Dipartimento di Fisica,
%Sapienza Universit\`a di Roma, P. le A. Moro 2, I-00185 Roma, Italy}

%\pacs{89.65.-s}{Social and economic systems}
%\pacs{89.75.Hc}{Networks and genealogical trees}
%\pacs{89.20.Hh}{World Wide Web, Internet}

\begin{abstract}
  The identification of which nodes are optimal seeds for spreading
  processes on a network is a non-trivial problem that has attracted much
  interest recently.  While activity has mostly focused on non-recurrent
  type of dynamics, here we consider the problem for the
  Susceptible-Infected-Susceptible (SIS) spreading model, where an outbreak
  seeded in one node can originate an infinite activity avalanche. We apply
  the theoretical framework for avalanches on networks proposed by Larremore
  {\em et al.}~[Phys. Rev. E \textbf{85}, 066131 (2012)], to obtain detailed
  quantitative predictions for the spreading influence of individual nodes
  (in terms of avalanche duration and avalanche size) both above and below
  the epidemic threshold.  When the approach is complemented with an
  annealed network approximation, we obtain fully analytical expressions for
  the observables of interest close to the transition, highlighting the role
  of degree centrality.  Comparison of these results with numerical
  simulations performed on synthetic networks with power-law degree
  distribution reveals in general a good agreement in the
  subcritical regime, leaving thus some questions
  open for further investigation relative to the supercritical
  region.
\end{abstract}

\maketitle

\section{Introduction}

Some epidemic outbreaks die out very rapidly, affecting just a few
individuals, while others, caused by the same pathogen, survive for much
longer times, infecting considerably larger fractions of the overall
population~\cite{complexspreading}. A meme posted by a celebrity on a social
network is rapidly spread to millions of users, while a similar meme posted
by a less connected individual rarely becomes viral.  What decides the fate
of a single spreading process?  One of the most important factors is where,
in the connectivity pattern mediating the spreading process, the initial
seed is located.  Intuitively, an infection seeded in a highly connected
node has a higher chance of reaching a large number of individuals, but
other, less local network features may in principle play a role: For
example, a node sitting at the boundary between two communities can be
highly effective in giving rise to large outbreaks even if having only a few
direct connections.  The non-trivial question of how to identify
``influential spreaders'' in a network has been the focus of an impressive
amount of work~\cite{Pei2014,Lu2016} in the last decade.  In a nutshell, the
problem is the prediction of how large is, on average, a spreading event
started by a single node in the network.\footnote{In the epidemic framework
a spreading event started in a single node is called an outbreak.  We will
often use in the following the more general term avalanche.}  A slightly
more limited goal is to rank network nodes depending on their spreading
capability and to correlate such a ranking with purely topological node
features, such as degree or other types of centrality.

The seminal paper by Kitsak et al.~\cite{Kitsak2010} posed in a clear manner
the question, highlighting its non-triviality and reporting a strong
correlation between the average size of an outbreak seeded in node $n$ and a
particular type of centrality measure, the $K$-core
index~\cite{Seidman1983269}. A huge number of other centralities have been
proposed and tested as ways to identify influential
spreaders~\cite{Lu2016,Chen2012,Li2014}, with different levels of success,
depending in many cases on the type of network substrate considered for the
empirical validation.  Among the most firmly grounded results,
Ref.~\cite{Radicchi2016} exploited a mapping to bond percolation to relate
spreading influence at the epidemic transition to nonbacktracking
centrality~\cite{hashimoto1989,Krzakala2013}, while Ref.~\cite{Min2018}
extended the approach presenting a message-passing framework allowing to
compute the average outbreak size for any value of the control parameter.

All these results were found for the Susceptible-Infected-Removed (SIR)
dynamics~\cite{PastorSatorras2015}, the simplest model for epidemics
allowing permanent immunity, or simple variants of it.  For this type of
models the definition of spreading influence is straightforward: Throughout
the whole phase-space, outbreaks have a finite duration and the average size
of an outbreak seeded in a node is the measure of how influential the node
is.  The other large class of epidemic models, whose simplest example is the
Susceptible-Infected-Susceptible (SIS) model, describes recurrent dynamics,
where each node can be reinfected over and over
again~\cite{PastorSatorras2015}. While some results are
available~\cite{Liu2017,Holme2018}, this class has received nevertheless
much less attention.  Our purpose in this paper is to fill this gap,
performing a systematic study of the problem of identifying influential
spreaders for a parallel, discrete-time,  version of SIS dynamics.  As for
non-recurrent (SIR-like) epidemic models, the phase diagram for SIS dynamics
consists in a healthy phase for small values of the control parameter (see
below) separated from an epidemic phase by a critical value of the control
parameter, the epidemic threshold.  While the nature of the healthy phase is
the same for both classes of models, the possibility of multiple
reinfections changes the nature of the epidemic phase. Above the threshold
there is a nonvanishing chance that a stationary ``endemic'' state is
reached, where the epidemic survives forever, with a steady finite density
of infected nodes.  This possibility implies that for SIS dynamics there are
different possible definitions of what an influential spreader is.  Below
the critical point, since all avalanches eventually end, one can take as a
measure of spreading influence of node $n$ the total duration or the total
size of an avalanche seeded in $n$, just as in the SIR case.  In the
supercritical region instead, some avalanches are infinite, i.e. they give
rise to the neverending (at least in the thermodynamic limit) endemic
state. In this case being an influential spreader may have two distinct
meanings.  Good spreaders can be nodes giving rise to an infinite avalanche
with large probability or nodes giving rise to large or long {\em finite}
avalanches.  These definitions of influential spreaders do not necessarily
coincide.  For this reason we consider three main observables of
interest~\cite{Larremore2012}: 

\begin{itemize}
\item
    The probability $b_n$ that an avalanche seeded in node $n$ is finite
    (i.e., it does not lead to the endemic state above the transition).
    Below the transition, $b_n=1$; above it, $b_n<1$.
\item
    The average duration $T_n$ of (finite) avalanches seeded in node $n$.
    In the following we will consider a discrete-time dynamics, hence the
    duration of each avalanche will be an integer.
\item 
    The average size $S_n$ of (finite) avalanches seeded in node $n$.  This
    quantity is equal to the total number of activation events.  Since each
    node can be activated more than once, $S$ can in principle be larger
    than the total number of nodes $N$.
\end{itemize}

Another observable of interest is the coverage, the average number of
distinct nodes reached by an avalanche~\cite{Boguna2013}. We performed some
comparisons between this quantity and $S_n$, finding minimal
differences. For this reason we will not consider the coverage in the
following.

\begin{comment}
The paper is structured as follows: In Section~\ref{sec:mean-field-theory}
we present analytical results for a parallel dynamics version we present we
summarize the results of the theoretical framework presented in
Ref.~\cite{Larremore2012}, useful for our analysis.  In addition, by means
of an annealed network approximation, we are able to perform calculations
explicitly, leading to detailed, fully analytical, predictions.  In Section
III we describe how numerical simulations are performed.  Sections IV and V
are devoted to the comparison between theory and numerics for values of
$\gamma$ below and above $5/2$, respectively.  A discussion of possible ways
to go beyond the shortcomings of the theoretical are presented in Section
VI. Finally, Section VII presents a summary of our work and some
perspectives.
\end{comment}

\section{Theoretical approach}
\label{sec:mean-field-theory}
\subsection{Quenched Mean-Field theory}
\label{sec:quenched-field-theory}

The identification of influential spreaders for SIS dynamics is greatly
helped by the application of the general quenched mean-field
(QMF)~\cite{Wang03,PVM_ToN_VirusSpread,Gomez10,Castellano2010} theoretical
approach for avalanches on networks proposed by Larremore et al. in
Ref.~\cite{Larremore2012}.  In that paper a comprehensive theory is
presented for the statistical properties of avalanches when the dynamical
process is determined by a matrix $A_{nm}$, whose individual entry is the
probability that the avalanche propagates from node $n$ to node $m$.  The
approach predicts a transition between a subcritical regime with all
avalanches finite ($b_n=1$) and a supercritical regime where a fraction
$1-b_n$ of all avalanches lasts forever.  The transition depends on the
largest eigenvalue $\lambda$ of the $A$ matrix.  If
$\lambda<\lambda_c^{QMF}=1$ the system is subcritical, if
$\lambda>\lambda_c^{QMF}$ the system is supercritical and infinite
avalanches start to appear.

We can apply this theory to the SIS dynamics by considering a parallel
discrete-time version of it, analogous to the Independent Cascade Model
(ICM) version of SIR dynamics~\cite{Goldenberg01}.  In this dynamics, at
each time step each infected node attempts to transmit the infection to all
its direct neighbors and each independent attempt succeeds with probability
$p$. After attempting to infect all its neighbors an infected node recovers
and switches back to the susceptible state with a recovery time fixed to 1.
The probability that an avalanche propagates from node $n$ to $m$ is
therefore $A_{nm} = p a_{nm}$, where $a_{nm}$ is the unweighted adjacency
matrix of the network.  Clearly, the largest eigenvalue of
$A$ is $\lambda=p \Lambda_M$, where $\Lambda_M$ is the largest eigenvalue of
the adjacency matrix $a$.  As a consequence the epidemic threshold is
$p_c^{QMF} = 1/\Lambda_M$.

\subsubsection{Probability of observing a finite avalanche}

Within QMF theory the probability $b_n$ of having a finite avalanche is
obtained by solving iteratively the equation
\begin{eqnarray}
  \label{bn}
  b_n &=& \prod_{m=1}^N [(1-A_{nm})+ A_{nm}b_m] 
 \nonumber \\
      &=&\prod_{m=1}^N [ 1 - p a_{nm} (1 - b_m)].
      \label{eq:b_n_qmf}
\end{eqnarray}
As shown in Ref.~\cite{Larremore2012} this equation admits a solution
$b_n=1$ for any $\lambda$ and for $\lambda>1$, i.e. in the supercritical
regime, another stable solution $b_n<1$.

\subsubsection{Average avalanche duration}
\label{sec:aver-aval-durat}

Let us consider $c_n(t)$ as the probability that an avalanche has a duration
smaller than or equal to $t$, with $c_n(0) = 0$, by definition. The
probability $c_n(t)$ fulfills the recurrence equation~\cite{Larremore2012}
\begin{equation}
    c_n(t + 1) = \prod_{m=1}^N \left[ (1-A_{nm}) + A_{nm} c_m(t) \right].
    \label{eq:c_n}
\end{equation}
In the limit $t\to\infty$, $c_n(t) \to b_n$, where $b_n = 1$ in the
subcritical phase, and $b_n < 1$ in the supercritical phase; hence
Eq.~\eqref{eq:b_n_qmf}. Consider $c_n(t) = b_n - f_n(t)$, with $f_n(0) =
b_n$. For sufficiently large $t$, $f_n(t)$ is small, so that we can
linearize Eq.~\eqref{eq:c_n} to obtain~\cite{Larremore2012}
\begin{equation}
  \label{linear}
  f_n(t) = \sum_m D_{nm}f_m(t-1),
\end{equation}
where for the SIS process
\begin{equation}
  \label{eq:25}
  D_{nm} = \frac{A_{nm} b_{n}}{(1 - A_{nm}) + A_{nm} b_m} = \frac{p b_na_{nm}}{1
  - p + p b_m},
\end{equation}
that, in the subcritical phase, where $b_n=1$, reduces to $D_{nm} = A_{nm}=p
a_{nm}$.  The avalanche duration distribution $p_n(t)$ is given by
\begin{equation}
%  \label{eq:22}
  p_n(t) = c_n(t) - c_n(t-1) = f_n(t-1) - f_n(t).
\end{equation}
Therefore, we can write the average avalanche duration $T_n$ as
\begin{eqnarray}
  \label{eq:23} \nonumber
  T_n  &=& \sum_{t=1}^\infty t p_n(t) = \sum_{t=1}^\infty t f_n(t-1) -
           \sum_{t=1}^\infty t f_n(t) \\ \nonumber
       &=& f_n(0) + \sum_{t=2}^\infty t f_n(t-1) -
           \sum_{t=1}^\infty t f_n(t) \\ \nonumber
       &=& b_n + \sum_{t=1}^\infty (t+1) f_n(t) - \sum_{t=1}^\infty t f_n(t) \\ 
       &=& b_n +
           \sum_{t=1}^\infty f_n(t) \equiv b_n + R_n.
\end{eqnarray}

If $f_n(t)$ decreases sufficiently fast, we can assume that the linearization of
$f_n(t)$ is valid for all times $t$, to obtain
\begin{eqnarray}
  \label{eq:24} \nonumber
  R_n &=& \sum_{t=1}^\infty f_n(t) = \sum_m D_{nm} \sum_{t=1}^\infty f_m(t-1)\\ \nonumber
      &=&
          \sum_m D_{nm} \left[ f_m(0) + \sum_{t=1}^\infty f_m(t) \right] \\
   &=&  \sum_m D_{nm} \left[b_m + R_m \right].
\end{eqnarray}

Combining the previous equations, we have
\begin{equation}
  \label{eq:3}
  T_n = b_n + \sum_m D_{nm}T_m.
\end{equation}
We can obtain an exact expression at the QMF level by inverting this linear
relationship, obtaining
\begin{equation}
  \label{eq:4}
  T = [I - D]^{-1}B,
\end{equation}
where $T = (T_1, \ldots, T_N)^T$, $B = (b_1, \ldots, b_N)^T$, and $I$ is
the identity matrix.

\subsubsection{Average avalanche size}
\label{sec:aver-aval-size}

From Ref.~\cite{Larremore2012},
the generating function of the avalanche size distribution $p_n(s)$,
\begin{equation}
  \label{eq:2}
  \phi_n(z) = \sum_s e^{-sz} p_n(s),
\end{equation}
fulfills the equation
\begin{equation}
  \label{eq:26}
  \phi_n(z) = e^{-z} \prod_m [1+ \sum_m (\phi_m(z)-1) H_{nm}],
\end{equation}
where
\begin{equation}
    H_{nm} = \frac{b_m p a_{nm}}{1- p + b_mp}
\end{equation}
in the SIS process.  
Taking logarithms of both sides of Eq.~(\ref{eq:26}), deriving
with respect to $z$ and setting $z=0$ one obtains
\begin{equation}
\label{eq:13bis}
\frac{\phi_n'(0)}{\phi_n(0)} = -1 + 
\sum_m \frac{H_{nm} \phi_m'(0)}{1+ (\phi_m(0)-1) H_{nm}}.
\end{equation}
From the definition of the generating function we have $\phi_n(0)=1$,
while $\phi_n'(0)$ is minus the average avalanche size,
\begin{equation}
\phi_n'(0) =  \left.\frac{d \phi_n(z)}{dz}\right|_{z=0}  =-\sum_s s p_n(s) =-S_n.
\end{equation}
Hence Eq.~(\ref{eq:13bis}) reads
\begin{equation}
  \label{eq:13}
  S_n = 1 + \sum_m H_{nm}S_m.
\end{equation}
Inverting this linear relationship in matrix format, we can write the QMF
solution
\begin{equation}
  \label{eq:28}
  S = [I - H]^{-1} \mathbf{1}_N
\end{equation}
where $S = (S_1, S_2, \ldots, S_N)^T$ and $\mathbf{1}_N$ is a column
vector of ones, of size $N$.
Notice that, contrary to the case of $T_n$, no long-time assumption
has been made to derive Eq.~(\ref{eq:13}).

It is  interesting to observe that, due to the similarity relation $H_{nm} =
\frac{b_m}{b_n} D_{nm}$~\cite{Larremore2012}, the average size and duration
at the QMF level are not independent, but related by the identity
\begin{equation}
    T_n = b_n S_n.
    \label{eq:similarity}
\end{equation}
Below the threshold, when $b_n = 1$, average time and duration are equal.
This fact can be interpreted in terms of avalanches that infect just a new
node in average every time step. On the contrary, in the supercritical
regime, $b_n<1$ in general, and thus $S_n > T_n$ in average.  
We remark also that, so far, we have not made any assumption on the value 
of $p$, hence these predictions are valid throughout the whole phase-diagram. 

\subsection{Annealed network approximation}
\label{sec:solution-gamma-}

Eqs.~\eqref{eq:b_n_qmf},~(\ref{eq:3}) and~(\ref{eq:13}) constitute sets of
$N$ closed equations, whose straightforward solutions provide predictions
about the ability of each node in a generic network to spread influence.

It is possible to obtain fully analytical results (and thus obtain more
physical insight) by implementing an additional approximation.  The annealed
network approximation~\cite{Dorogovtsev2008} assumes that the
network is fully rewired at each time step while keeping individual degrees
fixed. In this way dynamical correlations among different nodes are
destroyed and the state of each node can depend only on its degree $k$.
Mathematically, the annealed network approximation consists in replacing,
for uncorrelated networks, the adjacency matrix by the probabilistic
form~\cite{Dorogovtsev2008}
\begin{equation}
  \label{eq:16}
  a_{nm} \simeq \frac{k_n k_m}{\av{k} N}.
\end{equation}
This allows us to write
\begin{equation}
  \label{eq:8}
  \sum_m a_{nm}F(k_m) \simeq k_n \frac{\av{kF}}{\av{k}}
\end{equation}
for any function $F(k)$ depending on the degree.

\subsubsection{Probability of observing a finite avalanche}

Taking logarithms on both sides of Eq.~(\ref{bn}), we have
\begin{eqnarray}
  \label{eq:17} \nonumber
  \ln b_n &=& \sum_{m=1}^N \ln \left[1 - p a_{nm} (1 - b_m) \right]\\ \nonumber
          &=& - p \sum_{m=1}^N a_{nm} (1 - b_m) 
          = - p k_{n} + p \sum_{m=1}^N
              a_{nm} b_m,
\end{eqnarray}
where we have assumed $p(1-b_m)$ to be small, i.e., either $p$ is small, or
the system is close to criticality ($b_n \simeq 1$).

Assuming that $b_n$ is a function of the degree $k_n$, and using
Eq.~(\ref{eq:8}), we can then write
\begin{equation}
  \label{eq:18}
  \ln b_n  = - p k_n\left( 1 - \frac{\av{b k}}{\av{k}} \right),
\end{equation}
leading to
\begin{equation}
  \label{eq:19} 
  b_n = %\exp\left[- k_n p \left( 1 - \frac{\av{b k}}{\av{k}}
  %\right) \right]
    %\equiv 
  e^{-k_n p \theta},
\end{equation}
where we have defined the quantity
\begin{equation}
  \theta =  1 - \frac{\av{bk}}{\av{k}},
\end{equation}
which depends implicitly on $p$, through the average $\av{bk}$. We can make
this dependence explicit by considering the self-consistent equation 
\begin{equation}
  \label{eq:25b}
  \theta = 1 - \sum_k
  \frac{k P(k)e^{-k  p \theta}}{\av{k}} \equiv \Psi(\theta).
\end{equation}
The value $\theta = 0$, corresponding to $b_n = 1$, is always a solution of
Eq.~\eqref{eq:25b}. The condition for the existence of an additional
non-zero solution is that the function $y=\Psi(\theta)$ crosses the line $y
= \theta$ at a point $\theta>0$, something that happens when 
$\left. \frac{d \Psi(\theta)}{d \theta}\right|_{\theta = 0} > 1$.  This condition
    translates into $p > p_c^\mathrm{ann}$, where the threshold
\begin{equation}
  \label{eq:7b}
  p_c^\mathrm{ann} = \frac{\av{k}}{\av{k^2}},
\end{equation}
corresponds to the inverse of the largest eigenvalue $\Lambda_M$ of the
adjacency matrix in the annealed network approximation.

For finite networks of size $N$, all moments of the degree distribution are
finite. In this case, we can expand $\Psi(\theta)$ to second order in
$p\theta$, corresponding again to the vicinity of the transition point,
\begin{equation}
%  \label{eq:19}
  \theta = 1  - \sum_k \frac{ k P(k)}{\av{k}} e^{-k p \theta} \simeq  p
  \theta \frac{\av{k^2}}{\av{k}} - \frac{\av{k^3}}{2 \av{k}} p^2 \theta^2.
\end{equation}
Solving the previous equation for non-zero $\theta$ we obtain
\begin{equation}
  \theta \simeq \frac{2 \av{k}}{p^2 \av{k^3}} \left[
  \frac{p}{p_c^\mathrm{ann}} - 1 \right] = \frac{2 \av{k}}{p^2 \av{k^3}}
  \left[\lambda_0  - 1 \right],
\end{equation}
where we have defined $\lambda_0 \equiv p/p_c^\mathrm{ann}$.  This leads to
the finite size expressions
\begin{equation}
  \label{eq:21}
%  b_n \simeq \exp \left\{  -k_n \frac{2 \av{k}}{ p \av{k^3}} \left[
%      \lambda_0 - 1 \right]  \right\}
  b_n =1-k_n \frac{2 \av{k}}{ p \av{k^3}} \left[
      \lambda_0 - 1 \right]
\end{equation}
and
\begin{equation}
  \label{eq:22}
  1 - \av{b} \equiv 1 - \frac{\sum_n b_n}{N} = \frac{2 \av{k}^2}{
    p \av{k^3}} \left[ \lambda_0 - 1 \right].
\end{equation}

Notice that the last three equations correspond to expansions close
  to the critical point, and for this reason they are valid only for
  $\lambda_0 - 1$ small, in particular in the regime where $2 \av{k} (\lambda_0
    -1) / [p \av{k^3}]$ is smaller than one.

\subsubsection{Average avalanche duration}

Close to and below the critical point, i.e., assuming again $p(1-b_m)
\ll 1$ we can write $D_{nm} \simeq p b_n a_{nm}$.  Plugging this into
Eq.~(\ref{eq:3}) we have
\begin{equation}
  \label{eq:5}
  T_n = b_n +  p b_n\sum_m a_{nm}T_m,
\end{equation}
that can be written as
\begin{equation}
  \label{eq:7}
  X_n = 1 + p \sum_{m} b_m a_{nm} X_m,
\end{equation}
where $X_n = T_n / b_n$. Given Eq.~(\ref{eq:7}), assuming that $X_n$ is a
function of $k_n$, and in view of Eq.~\eqref{eq:8}, we can make the ansatz
$X_n = 1 + {\tilde A} k_n$, with ${\tilde A}$ a constant to be determined. 
Inserting the ansatz into Eq.~(\ref{eq:7}), we have
\begin{equation}
  \label{eq:6}
  {\tilde A} k_n = p  \sum_{m} a_{nm} b_m + {\tilde A} p \sum_m a_{nm}k_m b_m.
\end{equation}

Using the annealed network approximation, Eq.~(\ref{eq:8}), we have
\begin{equation}
  \label{eq:9}
  {\tilde A} k_n = p k_n \frac{\av{kb}}{\av{k}} + 
{\tilde A} p k_n  \frac{\av{k^2b}}{\av{k}},
\end{equation}
leading to
\begin{equation}
  \label{eq:10}
  {\tilde A}  = \frac{ p \av{kb}}{\av{k} - p \av{k^2b}}
\end{equation}
so that
\begin{equation}
  \label{eq:11}
  T_n = b_n \left(1+ \frac{ p \av{kb}}{\av{k} - p \av{k^2b}}\right)
\end{equation}

In the subcritical phase, $b_n=1$ and we obtain the simple explicit form
\begin{equation}
  \label{eq:12}
T_n = 1 + k_n \frac{p \av{k}}{\av{k} - p\av{k^2}} =
1 + k_n \frac{p}{1 - \lambda_0}.
\end{equation}
In the supercritical phase, $T_n$ depends on moments of $k$ and $b$ in
a more complicated way. 
Assuming we are close to the critical point, in a finite network,
we can use Eq.~\eqref{eq:21} to obtain
\begin{eqnarray}
  \label{eq:31}
  \frac{\av{kb}}{\av{k}} &=& 1 - \frac{2 \av{k^2}}{p \av{k^3}} (\lambda_0 -1)\\
   \frac{\av{k^2b}}{\av{k}} &=& \frac{1}{p_c^\mathrm{ann}} -
                                \frac{2}{p}(\lambda_0 -1).
\end{eqnarray}
Substituting into Eq.~\eqref{eq:11}, we finally have
\begin{equation}
  \label{eq:1}
  T_n = b_n \left[1+  k_n\left( \frac{p}{\lambda_0 -1} -
    \frac{2\av{k^2}}{\av{k^3}}\right) \right],
\end{equation}
that can be written in the fully explicit form
\begin{eqnarray}
  \label{eq:1bis}
  T_n &=& \left[ 1-k_n \frac{2 (p\av{k^2}-\av{k})}{p\av{k^3}} \right] \nonumber \\
      &\times& \left[1+  k_n\left( \frac{p \av{k}}{p\av{k^2}-\av{k}} -
    \frac{2\av{k^2}}{\av{k^3}}\right) \right].
\end{eqnarray}

\subsubsection{Average avalanche size}

From Eq.~\eqref{eq:similarity}, we have $S_n = T_n/b_n$, which translates,
from the annealed network solution for $T_n$, into
\begin{equation}
  \label{eq:29}
  S_n = 1 +  k_n  \frac{ p \av{kb}}{\av{k} -
  p \av{k^2b}}.
\end{equation}
Hence in the subcritical phase we have
\begin{equation}
  \label{eq:15}
  S_n = T_n = 1 + k_n  \frac{p}{1 - \lambda_0},
\end{equation}
while in the supercritical phase
\begin{eqnarray}
  \label{eq:27}
  S_n &=& 1 + k_n\left( \frac{p}{\lambda_0 -1} -
    \frac{2\av{k^2}}{\av{k^3}}\right) \nonumber \\
      &=& 1 + k_n\left(\frac{p \av{k}}{p\av{k^2}-\av{k}} -
    \frac{2\av{k^2}}{\av{k^3}}\right).
\end{eqnarray}

The relation $S_n = T_n/b_n$ implies that in the supercritical case
avalanche size and duration are related in a nontrivial way.  The explicit
expressions point out that the average size (minus one) is proportional to
the degree of the seed close to the epidemic threshold. The same is true for
the average duration, but nonlinear effects are stronger in this case.

\section{Numerical simulations}
\begin{figure}
  \includegraphics[width=\columnwidth]{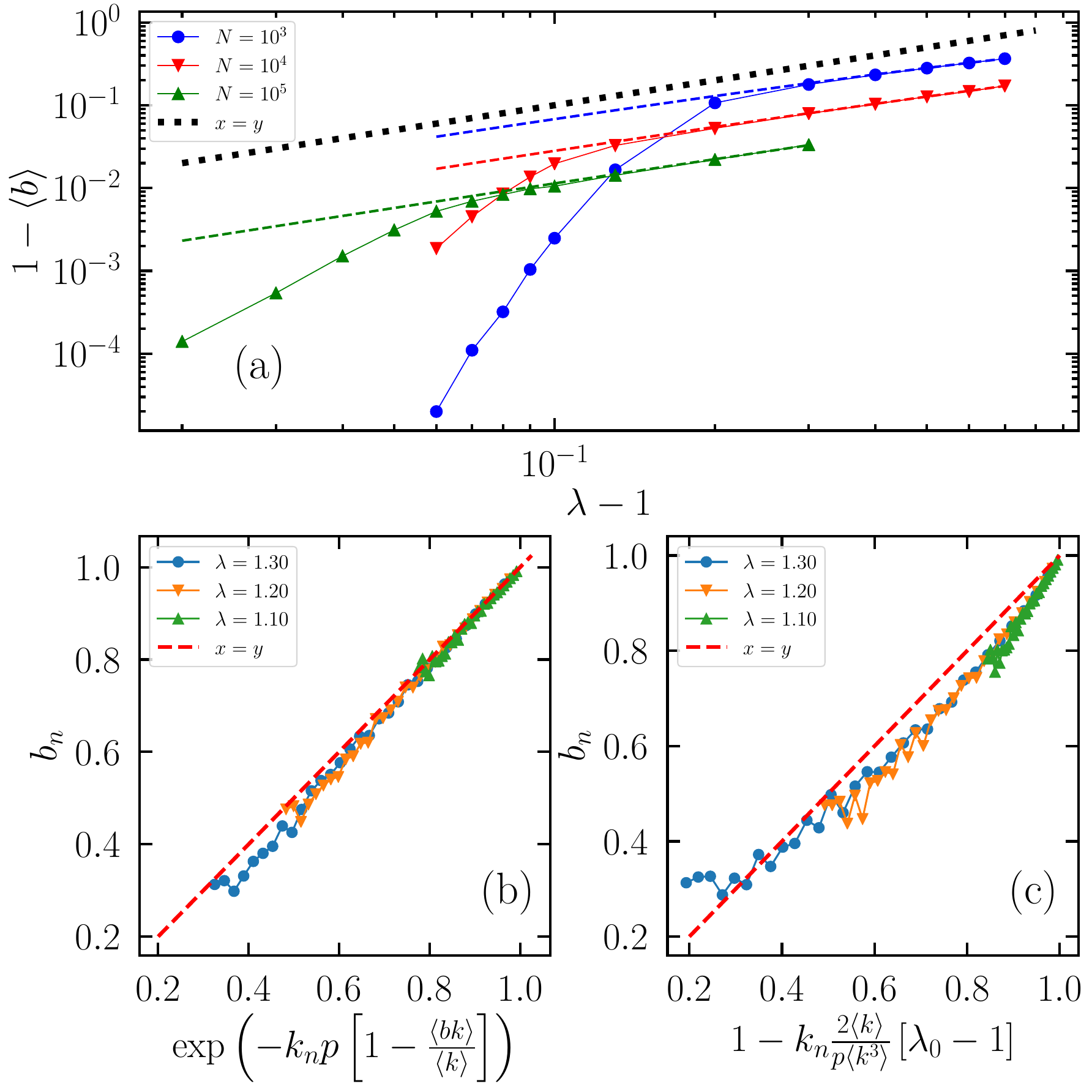}

  \caption{Comparison of $1-\av{b}$ vs $\lambda=p\Lambda_M$ from numerical
      simulations for $\gamma=2.25$ (symbols), with QMF theoretical
      predictions based on Eq.~(\ref{bn}) (dashed lines) for different
      values of $N$.  (b) and (c): Comparison of $b_n$ from numerical
      simulations for various $\lambda=p\Lambda_M$ and $N=10^4$ with
      annealed network predictions, Eqs.~(\ref{eq:19}) and~(\ref{eq:21}),
      respectively. In (c), $\lambda_0 = p \av{k^2}/\av{k}$ corresponds to
      the annealed network threshold.}
 \label{fig:bn_small}
\end{figure}

In this Section we perform a systematic comparison between the results of
the theoretical approaches presented in the previous Section and numerical
results obtained by simulating the SIS dynamics on synthetic uncorrelated
power-law distributed networks, built according to the Uncorrelated
Configuration Model~\cite{Catanzaro2005}.  The comparison is aimed at
validating the accuracy of the theoretical predictions and at identifying
the origin of possible inaccuracies in the various approximations performed.
The networks considered have a degree distribution $P(k) \sim k^{-\gamma}$
between a minimum value $k_{min}=3$ and a maximum $k_{max}=N^{1/2}$ (for
$\gamma<3$) and $k_{max}=N^{1/(\gamma-1)}$ for $\gamma>3$.

We consider as infinite all avalanches whose duration is longer than
$t_{inf}=300$ steps.  This threshold, chosen for computational
convenience, is larger than the average duration of finite avalanches except
for a narrow interval around the critical point.  
To improve the readability of figures, plotted values of $T_n$, $S_n$ and 
$b_n$ are binned over suitably chosen intervals.

We consider two values of $\gamma$ corresponding to the two different
scenarios occurring for the SIS transition~\cite{PastorSatorras2015}.  For
$\gamma<5/2$ the transition is triggered by a subextensive network subset
(the max K-core), composed of densely mutually interconnected
nodes~\cite{Castellano2012}.  In this case QMF theory and the annealed
network approximation are known to work well, providing an estimate of the
position of the epidemic threshold which is asymptotically exact in the
large size limit~\cite{Ferreira2012,Silva2019}.  For $\gamma>5/2$ the
epidemic transition is triggered by a different and highly nontrivial
mechanism.  Nodes with high connectivity (large hubs) considered together
with their immediate neighbors, form star graphs.
When $p$ becomes larger than $p_c^{QMF}=1/\Lambda_M \approx 1/\sqrt{\kmax}$
the largest of these star graphs becomes able to sustain epidemic activity
for a long time, proportional to $\exp(\kmax)$.  For larger values of $p$
other star graphs can independently sustain activity.  QMF theory describes
only the behavior of isolated stars.  However, the epidemic in each of them
survives for a time proportional to $\exp(k_i)$, i.e., there is no endemic
activity in the system (surviving for a time $\exp(N)$) if stars are
isolated~\cite{Goltsev2012,Lee2013}.  As a consequence the QMF estimate
$p_c^{QMF}=1/\Lambda_M$ is only a (not strict) lower-bound for the actual
epidemic threshold.  The consideration of long-range interactions among
stars, neglected by QMF theory, allows to understand how endemic activity
can be established and to calculate the actual critical threshold 
$p_c > p_c^{QMF}$~\cite{Boguna2013,Castellano2020}.

\begin{figure}
  \includegraphics[width=\columnwidth]{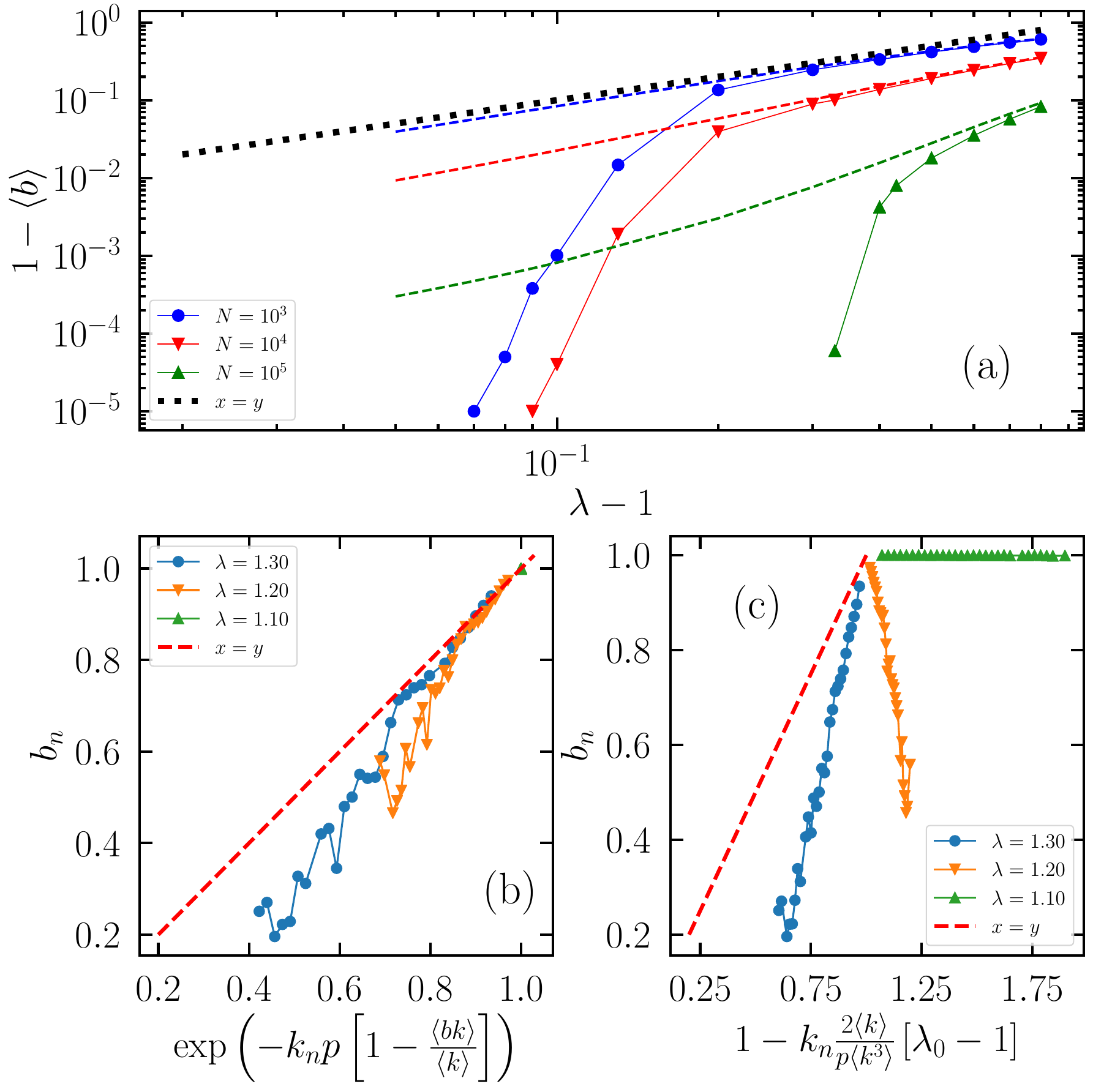}
  \caption{Comparison of $1-\av{b}$ vs $\lambda=p\Lambda_M$ from numerical
      simulations for $\gamma=3.50$ (symbols), with QMF theoretical
      predictions based on Eq.~(\ref{bn}) (dashed lines) for different
      values of $N$.  (b) and (c): Comparison of $b_n$ from numerical
      simulations for various $\lambda=p\Lambda_M$ and $N=10^4$ with
      annealed network predictions, Eqs.~(\ref{eq:19}) and~(\ref{eq:21}),
  respectively. In (c), $\lambda_0 = p \av{k^2}/\av{k}$ corresponds to the
  annealed network threshold.}
  \label{fig:bn_large}
\end{figure}

\setcounter{subsubsection}{0}

\subsubsection{Probability of observing a finite avalanche}

In Fig.~\ref{fig:bn_small}(a) we first compare the probability $1-\av{b}$ of
observing a finite avalanche calculated within the QMF approach
[Eq.~(\ref{bn})] with numerical results for $\gamma=2.25$. It turns out
clearly that QMF is very accurate in reproducing the behavior of $1-\av{b}$. 
The bending of the numerical curves
for small $\lambda-1$ is due to finite size effects reflecting the
existence of a size-dependent effective threshold.
As $N$ increases the
agreement between theory and numerics extends to lower values of
$\lambda-1$, confirming that the QMF threshold estimate is asymptotically
exact~\cite{Ferreira2012,Silva2019}.  The other two panels display instead
the accuracy of the fully analytical predictions obtained when also the
annealed network approximation is implemented.  Also in this case we find
a very good agreement between theory and simulations. Note that no parameter
has been fitted.

In Fig.~\ref{fig:bn_large} the same analysis is performed for $\gamma=3.50$.
The scenario is completely different.  Also in this case the agreement
between theory and simulations for what concerns $1-\av{b}$ holds above
a size-dependent threshold, but now the threshold moves toward {\em larger}
values of $\lambda-1$ as $N$ is increased. This shows that the actual SIS
threshold is larger than the QMF prediction, $\lambda_c > \lambda_c^{QMF}=1$,
and the discrepancy grows with the system
size~\cite{Boguna2013,Castellano2020}.
  The disagreement is even more
  evident in panels (b) and (c).
  In panel (b), all $b_n=1$  for $\lambda=1.1$ despite the QMF
  prediction that $b_n<1$ for $\lambda>\lambda_c^{QMF}=1$.
  This is a consequence of the fact that the true value of $\lambda_c$
  is larger than 1.1, so that $\lambda=1.1$ is actually subcritical
  instead of supercritical.
  In panel (c) we have the additional complication that, when
  $\lambda=p\Lambda_M$ is 1.1 or 1.2, $\lambda_0=p\av{k^2}/\av{k}$ is
  smaller than 1. This gives rise to the unphysical values larger than 1
  along the horizontal axis in Fig.~\ref{fig:bn_large}(c).
\begin{figure}
  \includegraphics[width=\columnwidth]{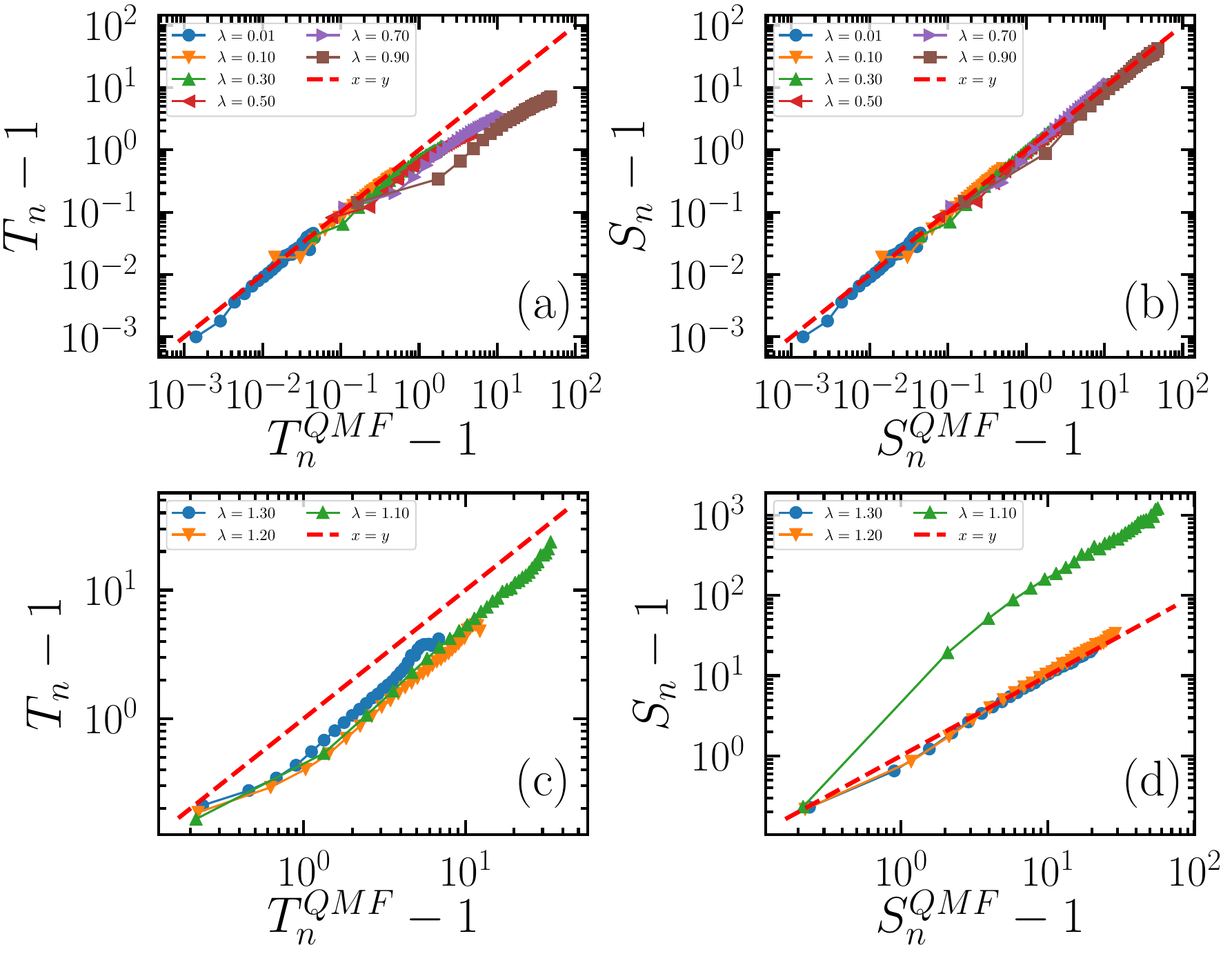}
  \caption{Comparison of numerical simulations with QMF theoretical
      predictions (Eqs.~(\ref{eq:4}), and~(\ref{eq:28})) for $\gamma=2.25$
      and $N=10^4$. (a) $T_n$ subcritical; (b) $S_n$ subcritical; (c) $T_n$
  supercritical; (d) $S_n$ supercritical.}
  \label{fig:exact_small}
\end{figure}

In summary,
for $\gamma <5/2$ both QMF and annealed network theories predictions
for the threshold are asymptotically exact, so that one expects
theory and simulations to perfectly agree for large $N$.
For $\gamma>5/2$ instead both theories predict a threshold value
not matching numerics as $N$ grows. In that case the predictions
for $b_n$ and $\av{b}$ work only sufficiently inside the supercritical
phase.

\subsubsection{Average avalanche duration and size}

In Fig.~\ref{fig:exact_small} we report a comparison between QMF predictions
(along the horizontal axis) and numerical results (vertical axis)
for $\gamma=2.25$. 
Results are presented for both the average duration $T_n$ (left column) and
the average avalanche size $S_n$ (right column). In the subcritical case
an excellent agreement is found for $S_n$ for any value of $\lambda$,
while the same is true only for durations
up to $T_n \approx 1$, corresponding to small values of $\lambda$.
For larger durations a disagreement starts to appear for $T_n$. 
This difference can be rationalized based on 
the different treatments of Section~\ref{sec:quenched-field-theory}.
The prediction for $S_n$ is valid for any size, without any small size
assumptions.
This failure of QMF theory in predicting the
avalanche duration can be instead attributed to the breakdown of the
assumption made to obtain Eq.~(\ref{eq:24}). 
Close to the transition point, critical slowing down leads to a 
slow decay of $f_n(t)$ toward zero and it is not
appropriate to assume linearization to hold starting from $t=1$.

For the average avalanche size in the supercritical case, 
Fig.~\ref{fig:exact_small}(d), large discrepancies between theory 
and simulations occur for $\lambda$ slightly larger than 1, 
while an excellent agreement is found again for larger $\lambda$.
This finding can be ascribed to the intrinsic impossibility
of discriminating precisely finite or infinite avalanches 
in networks of finite size.
At the critical point the distribution of avalanche times
features a power-law decay plus incipient infinite avalanches,
characterized by long but still finite durations. Some of them
are classified as finite according to the criterion $T < t_{inf}$
and therefore contribute to $T_n$ increasing its value by a large amount.
Reducing $t_{inf}$ alleviates this problem, but at the price of
risking to spuriously consider as infinite avalanches those belonging 
to the power-law decay.
This difficulty is naturally mitigated by considering larger systems,
or going to larger $\lambda$,
as the separation between the two components of the duration distribution
becomes more pronounced.
For the average avalanche duration in the supercritical case, 
Fig.~\ref{fig:exact_small}(c) this intrinsic problem with finite
size (which tends to increase $T_n$) competes with the breakdown
of the linearization for $f_n$ (which tends to decrease $T_n$).
As a consequence, numerical results are not far from QMF predictions.

\begin{figure}
  \includegraphics[width=\columnwidth]{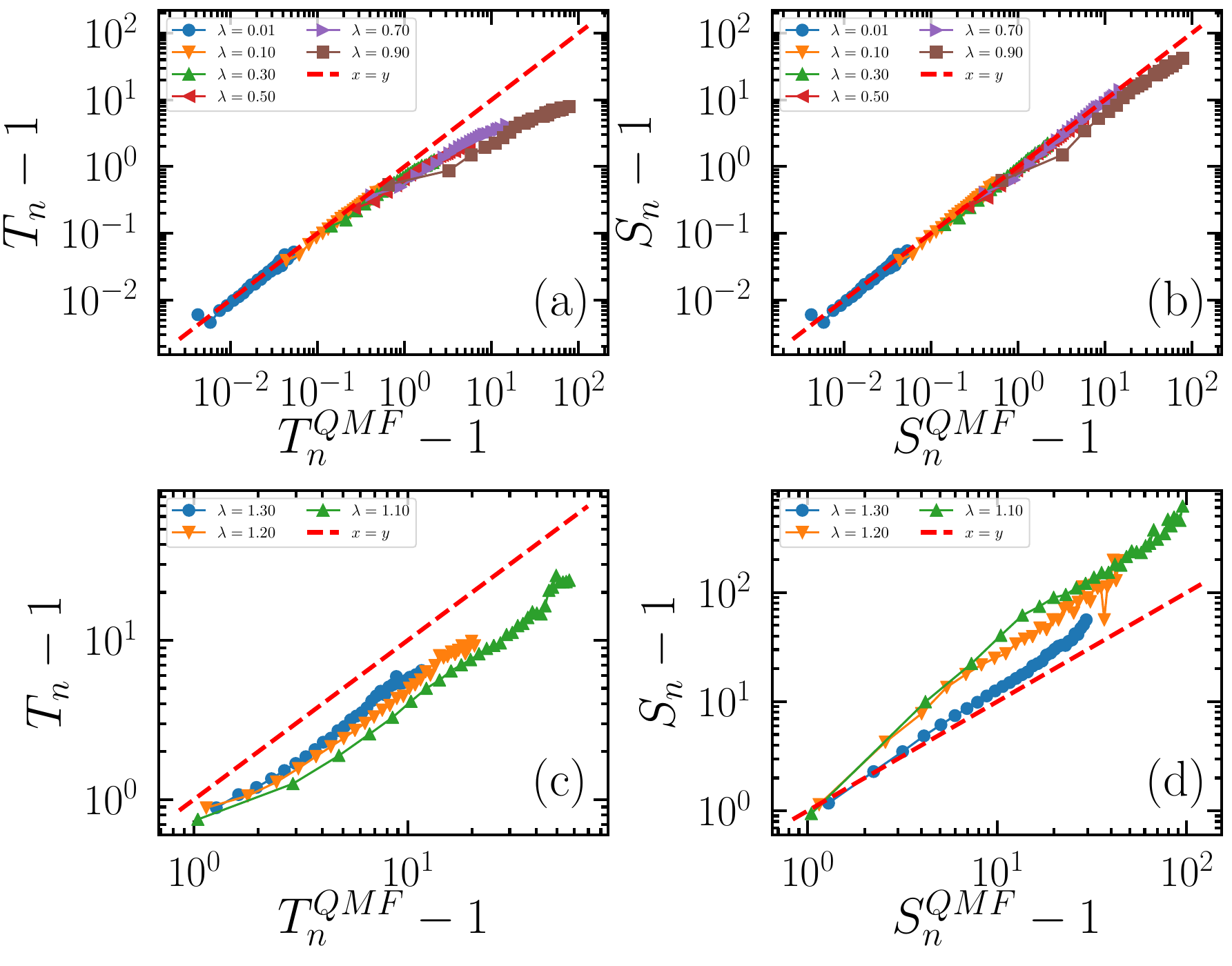}
  \caption{Comparison of numerical simulations with theoretical predictions
      (Eqs.~(\ref{eq:4}), and~(\ref{eq:28})) for $\gamma=3.50$ with
      $N=10^4$. (a) $T_n$ subcritical; (b) $S_n$ subcritical; (c) $T_n$
  supercritical; (d) $S_n$ supercritical.} \label{fig:exact_large}
\end{figure}
Figure~\ref{fig:exact_large} reports the same analysis for $\gamma=3.50$;
the same qualitative scenario emerges concerning the agreement between QMF
theory and simulations.  
This is somewhat surprising, given the known failure of QMF in describing
accurately the transition in this case. However, this failure becomes
most evident for very large networks, while only moderate values of $N$ are
considered here~\cite{Silva2019}.

\begin{figure}
  \includegraphics[width=\columnwidth]{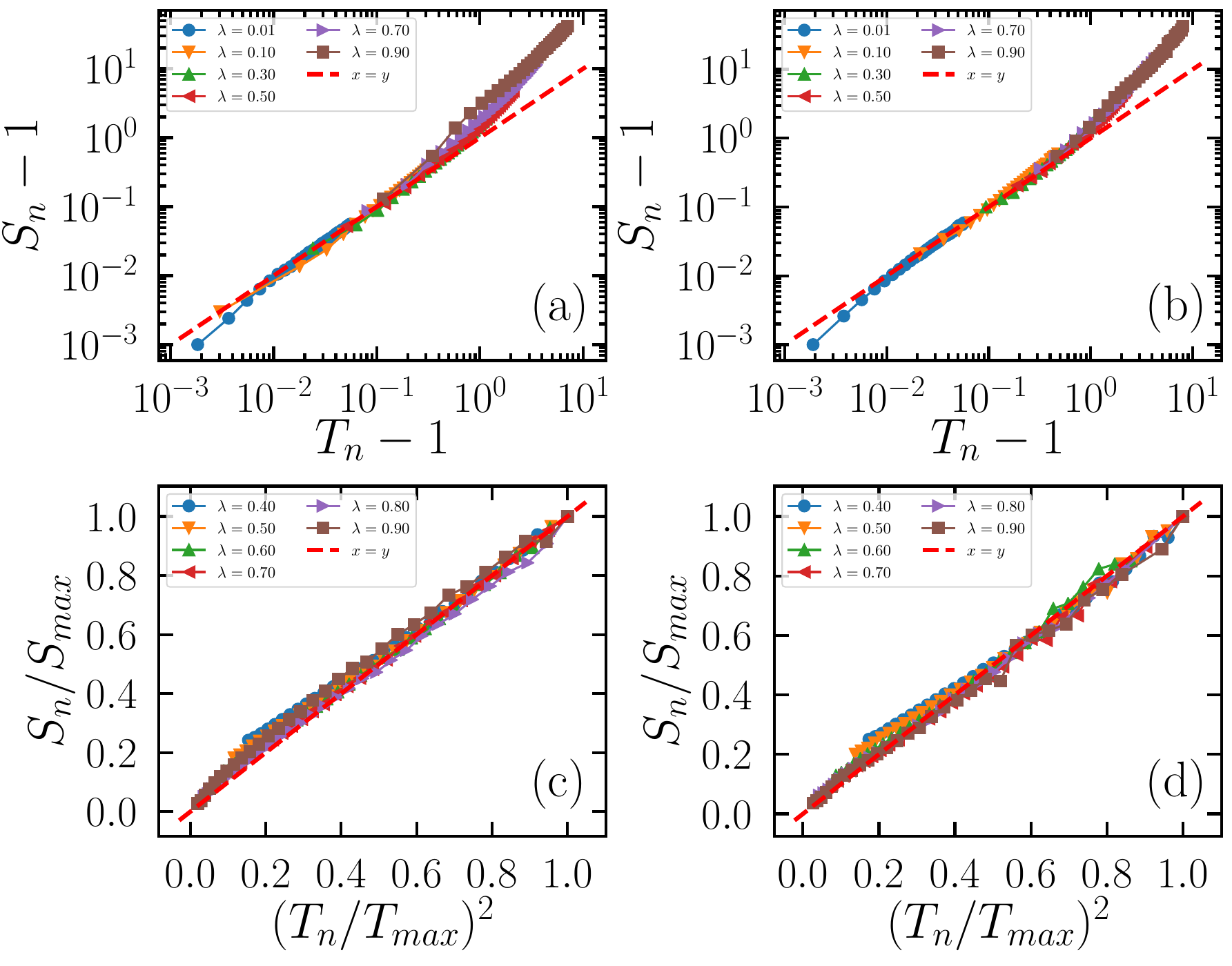} 
  \caption{Plot of the avalanche size $S_n$ as a function as the avalanche
      duration $T_n$ for $\gamma=2.25$ (a) and $\gamma=3.50$ (b) in the
      subcritical regime. Panels (c)
      and (d) show the avalanche size $S_n$ (normalized by its maximum
      value) as a function of the square of the avalanche duration  $T_n$
  normalized by its maximum value), for   $\gamma=2.25$ and $\gamma=3.50$,
  respectively. Data from numerical simulations in networks of size
  $N=10^4$.}
  \label{fig:sn_vs_tn}
\end{figure}

Further insight into what happens for long subcritical avalanches close to
the threshold is provided in Fig.~\ref{fig:sn_vs_tn}, where we plot the
average avalanche size versus the corresponding duration, obtained from
numerical simulations. As we can see here, for small values of $\lambda$,
the QMF prediction $S_n \sim T_n$ is well fulfilled. Closer to the critical
point, on the other hand, we observe a departure from the linear behavior,
which reflects the mismatch between the poor agreement with the QMF theory
for large $\lambda$ of the avalanche  duration $T_n$ and the very good 
agreement of the avalanche size $S_n$.
Numerically, we observe instead a good fit for $\lambda$ close to $1$ 
to the form $S_n \sim T_n^2$, see Fig.~\ref{fig:sn_vs_tn}(c) and~(d),
which is a consequence of the critical scalings of $p_n(s)$ and 
$p_n(t)$~\cite{Larremore2012}.

\begin{figure}
  \includegraphics[width=\columnwidth]{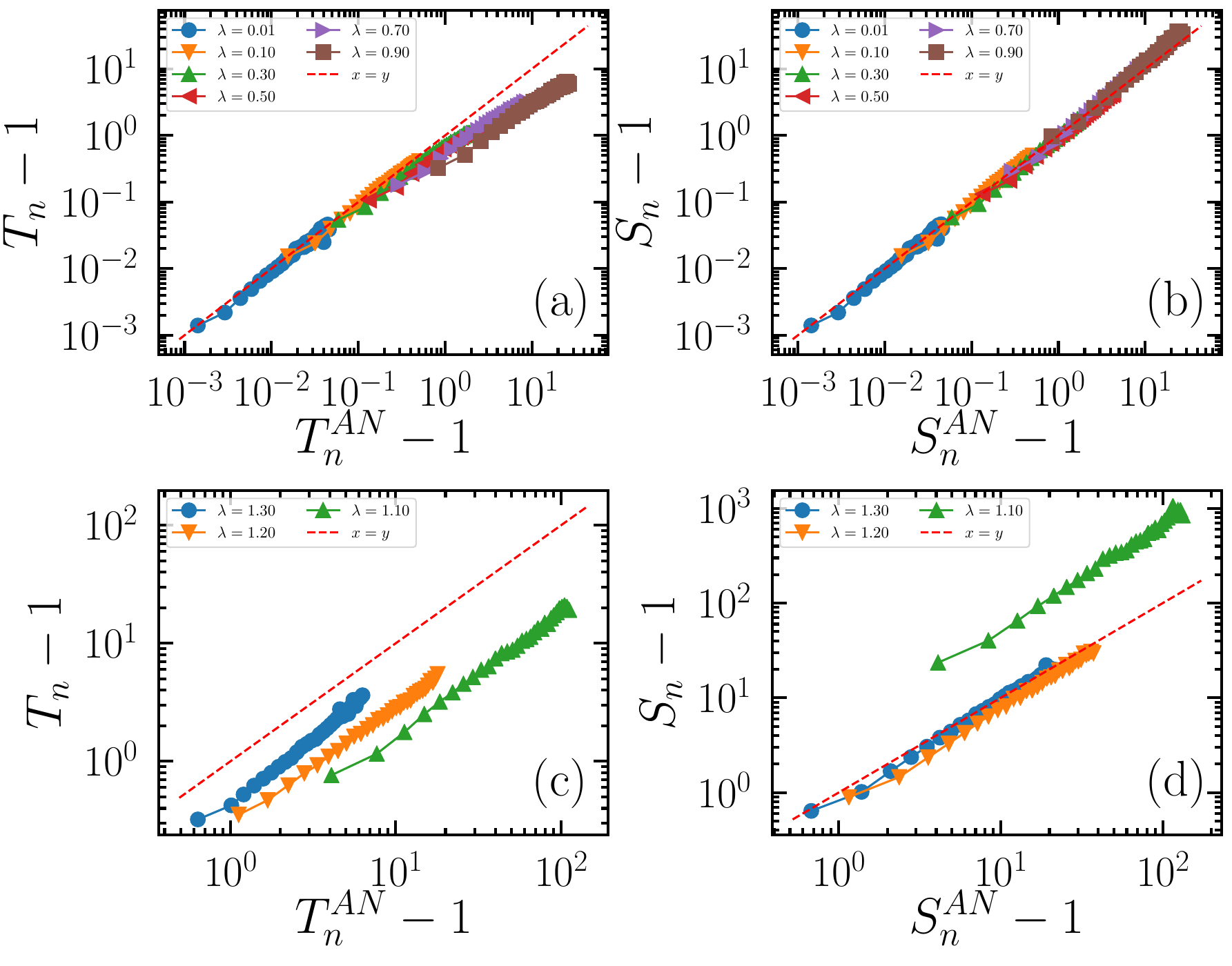} \caption{Comparison
      of numerical simulations with annealed network predictions for
      $\gamma=2.25$ and $N=10^4$. (a) $T_n$ subcritical (Eq.~(\ref{eq:12}));
      (b) $S_n$ subcritical (Eq.~(\ref{eq:15})); (c) $T_n$ supercritical
  (Eq.~(\ref{eq:1bis})); (d) $S_n$ supercritical (Eq.~(\ref{eq:27})).}
  \label{fig:an_small}
\end{figure}

\begin{figure}
  \includegraphics[width=\columnwidth]{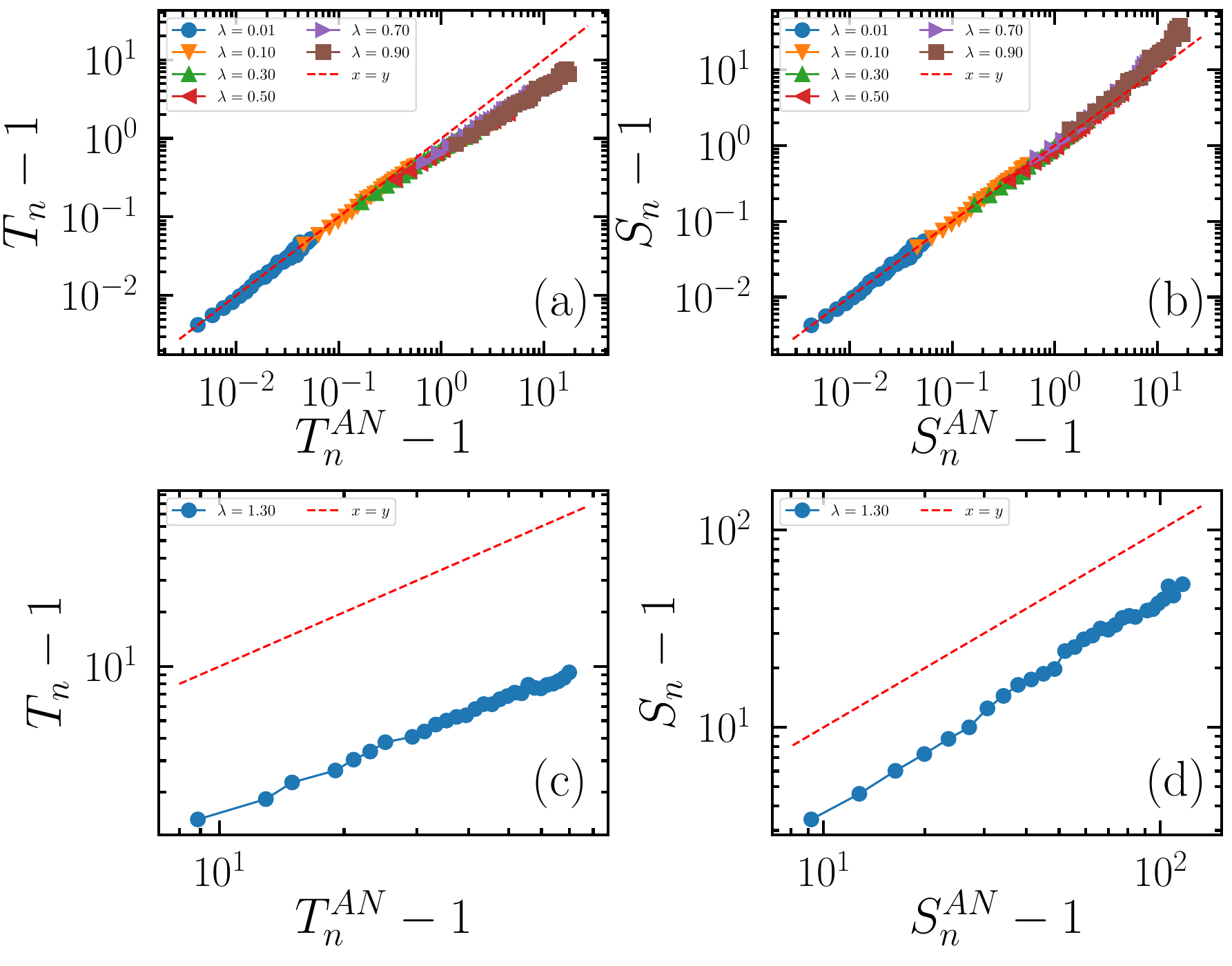} \caption{Comparison
      of numerical simulations with annealed network predictions for
      $\gamma=3.50$ and $N=10^4$. (a) $T_n$ subcritical (Eq.~(\ref{eq:12}));
      (b) $S_n$ subcritical (Eq.~(\ref{eq:15})); (c) $T_n$ supercritical
  (Eq.~(\ref{eq:1bis})); (d) $S_n$ supercritical (Eq.~(\ref{eq:27})).}
  \label{fig:an_large}
\end{figure}

Figures~\ref{fig:an_small} and~\ref{fig:an_large} analyze instead the
agreement between simulations and the theoretical results obtained using the
annealed network approximation, for both $T_n$ [Eqs.~(\ref{eq:12})
and~(\ref{eq:1bis})] and $S_n$ [Eqs.~(\ref{eq:15}) and~(\ref{eq:27})].
For $\gamma=2.25$ we find a very good agreement throughout the whole
phase-diagram, except for a critical region around the epidemic threshold.
For $\gamma=3.50$, the fit is very good in the subcritical regime, and 
rather poor away from the critical point in the supercritical regime. This is
due to the fact that the annealed network threshold $p_c^{ann}$ largely
overestimates the actual threshold.
A consequence of this overestimate is that for $\lambda$ slightly
larger than 1, Eqs.~\eqref{eq:1bis} and~\eqref{eq:27}
predict negative values of $T_n$ and $S_n$.
A better agreement between theory and numerical results is expected
  for larger values of $\lambda$, which will be less affected by the
mismatch between the prediction of the threshold and its actual value.

\begin{figure}
    \includegraphics[width=\columnwidth]{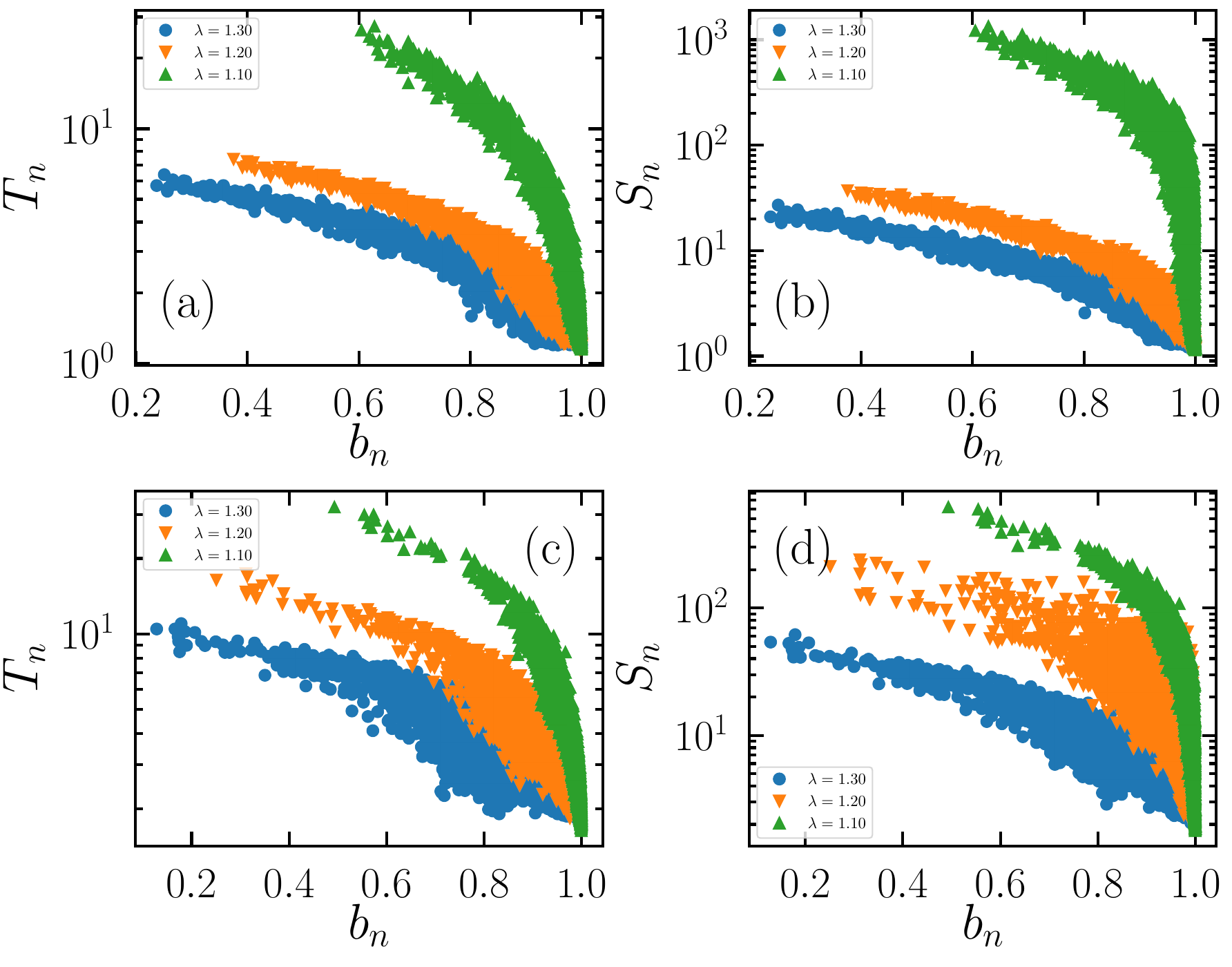}
    \caption{Dependence of the avalanche average duration and size as a
    function of the $b_n$ in the supercritical regime. (a) $T_n$ and (b)
$S_n$ for $\gamma=2.25$. (c) $T_n$ and (d) $S_n$ for $\gamma=3.50$.}
\label{fig:s_t_vs_b}
\end{figure}

Finally, in Fig.~\ref{fig:s_t_vs_b} we report the dependence of the average
avalanche duration $T_n$ and size $S_n$ in the supercritical regime, as a
function of the probability of observing a finite avalanche $b_n$.  For both
$\gamma=2.25$ and $\gamma=3.50$, we find continuously decreasing functions
of $b_n$, with large but not huge fluctuations.  The monotonicity allows us
to assume, as a rule of thumb, that if a node has a larger probability
$1-b_n$ to originate an infinite avalanche than a second node, the first
will also give rise on average to longer and larger {\em finite} avalanches.
In other words, ranking nodes according to $b_n$, $T_n$ or $S_n$ produces
the same ordering, apart from fluctuations.

\section{Conclusions}

In this paper we present a thorough investigation, both analytical and
numerical, of the problem of identifying influential spreaders for SIS
dynamics on generic networks.  We first apply to this problem the
theoretical QMF framework introduced by Larremore et al. in
Ref.~\cite{Larremore2012}. In this way we are able to write down closed
equations, whose numerical solution allows us to calculate all the
observables of interest in the whole phase-diagram.  Within this approach a
further step, consisting in the implementation of the annealed network
approximation, allows us to derive explicit analytical predictions valid
below and above (but close to) the critical point.  These predictions point
out the importance of degree centrality in determining the spreading
influence.  Comparison of these results with analytical simulations,
performed on synthetic networks with power-law distributed degrees $P(k)
\sim k^{-\gamma}$, confirms the substantial accuracy of the theoretical
approaches below the threshold, in particular for what concerns the
avalanche size. Above the expected QMF threshold discrepancies arise, which
tend to disappear inside the active region of the phase-diagram.

Let us discuss in detail strengths and limitations of the theoretical
approaches considered here.  The QMF theory of Larremore et
al.~\cite{Larremore2012} is based on three approximations: The neglect of
dynamical correlations among neighbors' states; the locally tree-like
assumption that allows to write down Eqs.~(\ref{bn}) and~(\ref{eq:c_n}); the
linearization entailed in Eq.~(\ref{linear}) for the derivation of $T_n$.
The annealed network approximation adds another assumption on top of them.
For strongly heterogeneous unclustered networks, such as those generated by
the Uncorrelated Configuration Model with $\gamma<5/2$, the first two
assumptions are essentially correct.  Studies on SIS dynamics have shown
that even the annealed network approximation works very well in this case.
The linearization holds instead only if one is sufficiently far from the
epidemic threshold. Close to it, the regime with exponential decay into the
stationary state is preceded by a long transient during which the decay is
power-law (critical slowing down), $f_n$ is not small and linearization does
not hold; the dominating contribution to the overall duration of avalanches
comes from this long preasymptotic regime.  For this reason predictions for
$T_n$ are not consistent with simulations, for values of $p$ close to the
threshold.  When $\gamma>5/2$, the agreement between our theoretical results
and simulations is expected to be poorer.  The reason is that already the
first approximation, the neglect of dynamical correlations, fails to capture
the complex physical mechanisms underlying the epidemic transition and thus
introduces large errors.  See Ref.~\cite{PastorSatorras2015} for more on
this.  A fundamental consequence is that already the QMF estimate of the
epidemic threshold is qualitatively and quantitatively incorrect.  This
leads to the expectation that predictions for influential spreaders in the
supercritical case are largely off target.  The relatively small errors
observed here are believed to be an effect of the relatively small network
sizes considered.  The additional annealed network approximation, which does
not work for $\gamma>5/2$, makes the theoretical approaches perform even
worse.  Recent results~\cite{Castellano2020} have pointed out the great
complexity of the physical mechanisms underlying the epidemic transition in
random networks with $\gamma>5/2$.  This understanding constitutes the
necessary groundwork for reaching a satisfactory predictive ability for
influential spreaders also in this case, a goal that remains to be reached.
Similarly, the extension of this approach to networks exhibiting
  more complicated features, such as clustering, correlations or
  mesoscopic structures, remains an interesting open issue.

Finally, it is worth remarking that our results have been derived for a
particular, discrete-time, version of SIS dynamics, rather
different from the usual continuous-time version. The present version,
with parallel dynamics and recovery time fixed to 1, allows a freshly
infected node to immediately reinfect the node that infected it in the 
preceding time step. This kind of process is instead strongly hindered
in continuous-time SIS, where the neighbor $j$
that infected node $i$ remains infected for a while so that the
number of connections of $i$ available for further spreading is 
effectively $k_i-1$.
This kind of dynamical correlations has strong effects for 
influential spreaders in SIR dynamics.  In that case degree centrality is a
good approximation of the spreading influence of individual nodes, but close
to the transition a better approximation is provided by the non-backtracking
centrality~\cite{Martin2014,Radicchi2016,Min2018}, which may differ markedly
from the former. 
The quest for accurate predictors of spreading influence for
continuous-time SIS dynamics close to the transition is an interesting
open avenue for further research.

\begin{acknowledgments}
  We acknowledge financial support from the Spanish Government's MINECO,
  under project FIS2016-76830-C2-1-P.  R. P.-S. acknowledges additional
  financial support from ICREA Academia, funded by the \textit{Generalitat
  de Catalunya} regional authorities.
\end{acknowledgments}

%\bibliography{Resubmit_Clean}

\end{document}